\documentclass[a4paper,fleqn,usenatbib]{mnras}
\usepackage{mathptmx}
\usepackage[T1]{fontenc}
\usepackage{ae,aecompl}
\usepackage{times}
\usepackage{amssymb}
\usepackage{amsmath}
\usepackage{upgreek}
\usepackage{color}
\usepackage{graphicx}

\title[W49A and W51 star formation]{Extreme star formation in the Milky Way: Luminosity distributions of young stellar objects in W49A and W51}
\author[D. J. Eden et al.]{D. J. Eden,$^{1}$\thanks{E-mail: D.J.Eden@ljmu.ac.uk} T. J. T. Moore,$^{1}$ J.S. Urquhart,$^{2}$ D. Elia,$^{3}$ R. Plume,$^{4}$ C. K\"{o}nig,$^{5}$ \newauthor A. Baldeschi,$^{3}$ E. Schisano,$^{3}$  A.J. Rigby,$^{6}$ L.K. Morgan,$^{1}$ M.A. Thompson$^{7}$\\
$^{1}$Astrophysics Research Institute, Liverpool John Moores University, IC2, Liverpool Science Park, 146 Brownlow Hill, Liverpool, L3 5RF, UK\\
$^{2}$School of Physical Sciences, Ingram Building, University of Kent, Canterbury, Kent CT2 7NH, UK\\
$^{3}$Istituto di Astrofisica e Planetologia Spaziali - INAF, Via Fosso del Cavaliere 100, I-00133 Roma, Italy\\
$^{4}$Department of Physics and Astronomy, University of Calgary, 2500 University Drive NW, Calgary, Alberta T2N 1N4, Canada\\
$^{5}$Max-Planck-Institut f\"{u}r Radioastronomie, Auf dem H\"{u}gel 69, 53121 Bonn, Germany\\
$^{6}$School of Physics and Astronomy, Cardiff University, Cardiff CF24 3AA, UK\\
$^{7}$Centre for Astrophysics Research, Science \& Technology Research Institute, University of Hertfordshire, College Lane, Hatfield, Herts AL10 9AB, UK}

\date{Accepted XXX. Received YYY; in original form ZZZ}

\pubyear{2018}

\begin{document}
\label{firstpage}
\pagerange{\pageref{firstpage}--\pageref{lastpage}}
\maketitle

\begin{abstract}

We have compared the star-formation properties of the W49A and W51 regions by using far-infrared data from the \emph{Herschel} infrared Galactic Plane Survey (Hi-GAL) and 850-$\upmu$m observations from the James Clerk Maxwell Telescope (JCMT) to obtain luminosities and masses, respectively, of associated compact sources. The former are infrared luminosities from the catalogue of \citet{Elia17}, while the latter are from the JCMT Plane survey source catalogue as well as measurements from new data. The clump-mass distributions of the two regions are found to be consistent with each other, as are the clump-formation efficiency and star-formation efficiency analogues. However, the frequency distributions of the luminosities of the young stellar objects are significantly different.  While the luminosity distribution in W51 is consistent with Galaxy-wide samples, that of W49A is top-heavy. The differences are not dramatic, and are concentrated in the central regions of W49A. However, they suggest that physical conditions there, which are comparable in part to those in extragalactic starbursts, are significantly affecting the star-formation properties or evolution of the dense clumps in the region.

\end{abstract}

\begin{keywords}

stars: formation -- ISM: individual objects: W49A -- ISM: individual objects: W51

\end{keywords}

\section{Introduction}

Two of the major open questions in star formation research are: what is the dominant mechanism regulating the efficiency and rate of star formation and on what scale does this mechanism operate. Increases in the average efficiency and rate of star formation are observed over large systems, i.e. starburst galaxies (e.g. \citealt{Scoville00}; \citealt{Dopita02}; \citealt{Kennicutt12}) and on the smaller scale of individual molecular clouds \citep[e.g.][]{Moore07, Polychroni12}.

Recent studies have attempted to determine the effect that the spiral arms, and other features of large-scale structure, have had on the efficiency of star formation in the Milky Way (\citealt{Eden12,Eden13,Eden15,Moore12,Ragan16,Urquhart17}). On average, the efficiencies were found to be roughly constant over kiloparsec scales, regardless of environment, with some minor enhancements associated with some, but not all, spiral arms.  Closer inspection showed that individual, extreme star-forming regions, namely the W49A and W51 complexes, were responsible for localised peaks in the ratio of infrared luminosity to molecular gas mass, even averaged over large sections of a spiral arm \citep{Moore12}. The study of \citet{Moore12} found that the star formation rate density ($\Sigma_{\rmn{SFR}}$ in units of M$_{\sun}$\,yr$^{-1}$\,kpc$^{-2}$) had significant increases at Galactocentric radii associated with spiral arms, but the vast majority of these increases, $\sim$ 70 per cent, were due to source crowding. The remaining 30 per cent of this increase was found to be due to the inclusion of these individual high-SFR star-forming regions. In the Sagittarius arm, thought to include W51, the increase was to be due to an increase in the number of young stellar objects (YSOs) per unit mass, whilst the increase seen towards the Perseus arm is thought to be due to the presence of W49A, which has a larger luminosity per YSO, i.e. the luminosity distribution in this region is flatter.

A change in the luminosity distribution of the stars in the W49A star-formation region would indicate a possible change in the stellar initial mass function (IMF). This would be very significant as a review of the IMF in environments from local clusters to nearby galaxies to starburst galaxies has found strong variations from the Salpeter-like form can be ruled out \citep{Bastian10}. As inferred, the IMF was found to be fairly constant within the Milky Way \citep{McKee07} but hints at variations have been detected in the extreme star-forming conditions within the Galactic Centre. These clusters have been shown to have significant variations in the IMF \citep{Espinoza09} but more recent observations indicate it is Salpeter-like \citep{Lockmann10,Habibi13}. However, a change in the W49A mass function compared to other significant star-forming regions (W43 and W51) would indicate a real deviation from the global IMF of the Galaxy.

W49A is at a distance estimated to be 11.11$^{+ 0.79}_{- 0.69}$\,kpc \citep{Zhang13} and is one of the most extreme star-forming regions in the Galaxy \citep[e.g.][]{Galvan-Madrid13}. This region is considered extreme as it has many quantities consistent with those found in LIRGS and ULIRGS, (ultra)luminous infrared galaxies, with localised dust temperatures of $>$\,100\,K and column densities $>$\,10$^{5}$\,cm$^{-3}$ \citep{Nagy12} and a luminosity per unit mass of $\sim$ 10\,L$_{\sun}$/M$_{\sun}$, compared to $\sim$ 100\,L$_{\sun}$/M$_{\sun}$ in ULIRGS \citep[e.g.][]{Solomon97}. The absolute luminosity of W49A ($\sim$ 10$^{7}$\,L$_{\sun}$; \citealt*{Harvey77,Ward-Thompson90}) does not compare to those of LIRGS and ULIRGS ($\sim$\,10$^{11}$\,--\,10$^{12}$\,L$_{\sun}$), but a mass of $\sim$ 10$^{6}$\,M$_{\sun}$ \citep{Sievers91} gives a $L/M$ that is within an order of magnitude. The region also has an overabundance of ultra-compact \ion{H}{ii} regions, with a factor of $\sim$ 3 more found coincident with this region compared to any other in the first quadrant of the Galaxy \citep{Urquhart13}.

The star-forming region W51 has a comparable $L/M$ to W49A (2.3\,$\times$\,10$^{5}$\,M$_{\sun}$, 3\,$\times$\,10$^{6}$\,L$_{\sun}$; \citealt{Harvey86,Kang10}) and has starburst-like star formation, with the majority occurring recently \citep[e.g.][]{Clark09} and very efficiently \citep{Kumar04}. W51 is at an estimated distance of 5.41$^{+ 0.31}_{- 0.28}$\,kpc \citep{Sato10}. Distances to both W49A and W51 are from maser parallax measurements.

The aim of this paper is to determine the star-forming properties of the two regions, building on the work of \citet{Moore12}, who found that the presence of these two regions was producing significant increases in the mean $L/M$ on kiloparsec scales. Assuming that the IMF is fully sampled, invariant, and that the infrared-bright evolutionary stages have lifetimes short compared to those of molecular clouds, then we would expect $L/M$ to be correlated with SFE. Alternatively, changes in $L/M$ may be due to variations in the luminosity distribution of the embedded massive YSOs, suggesting variations in the IMF.

We use data from the James Clerk Maxwell Telescope (JCMT) Plane Survey (JPS; \citealt{Moore15,Eden17}), additional 850-$\upmu$m continuum data from the JCMT, and the \emph{Herschel} infrared Galactic Plane Survey \citep{Molinari10a, Molinari10b} to determine the distribution of clump masses and embedded YSO luminosities for both regions, and examine the relationship between luminosity and mass. 

The paper is structured as follows: Section 2 introduces the data, Section 3 describes how the sources are selected for the study as well as the methods used to calculate source mass and luminosity. Section 4 presents the results, with Section 5 discussing those results. In Section 6 we provide a summary of our results and give conclusions.

\section{Data}

\subsection{\emph{Herschel} infrared Galactic Plane Survey}

The \emph{Herschel}\footnote{{\em Herschel} was an ESA space observatory with science instruments provided by European-led Principal Investigator consortia and with important participation from NASA.} infrared Galactic Plane Survey (Hi-GAL; \citealt{Molinari10b,Molinari10a}) was an Open-time Key Project of the $\emph{Herschel Space Observatory}$, and has mapped the entire Galactic Plane, with the inner Galaxy portion and initial compact-source catalogues outlined by \citet{Molinari16a,Molinari16}. This section, spanning Galactic longitudes of $-$70$\degr$\,$\leq$\,$\emph{l}$\,$\leq$\,68$\degr$, contains the W49A and W51 star-forming regions, imaged with the PACS \citep{Poglitsch10} and SPIRE \citep{Griffin10} cameras at 70, 160, 250, 350 and 500\,$\upmu$m with diffraction-limited beams of 6--35 arcseconds \citep{Molinari16a}. The catalogue was produced using the source extraction algorithm, CuTEx (Curvature Threshold Extractor; \citealt{Molinari11}), with a band-merged catalogue produced by \citet{Elia17}.

The Hi-GAL data have saturated pixels present in all five wavebands within both the W49A and W51 regions \citep{Molinari16a}. These saturated pixels occur in the most central areas of the two regions. However, only W51 is significantly affected, with over 300 pixels in the 250-$\upmu$m data. The saturated regions in W49A are not associated with any significant dust clumps identified by ATLASGAL \citep{Urquhart14b}. Accounting for the saturation in the W51 region will be discussed in Section 5.2.

The Hi-GAL sources used in this study are compact objects, tracing the peaks of the luminosity found embedded within the larger, star-forming clump structures. The fixed-aperture-based photometry, which is described below and in full in \citet{Elia17}, may produce fluxes and luminosities that depend on this method.

\subsection{JCMT continuum data}

The two regions were imaged in the 850-$\upmu$m continuum by the Submillimetre Common-User Bolometer Array 2 (SCUBA-2) instrument \citep{Holland13} on the JCMT at an angular resolution of 14.4\,arcseconds. The W51 data are taken from the JCMT Plane Survey (JPS: \citealp{Moore15})\footnote{The JPS is part of the JCMT Legacy Surveys Project \citep{Chrysostomou10}.} where the compact sources are catalogued in \citet{Eden17}. The W49A data were obtained in standard time allocations under Project IDs m13bu27 and m14au23.

\begin{figure*}
\includegraphics[width=0.99\linewidth]{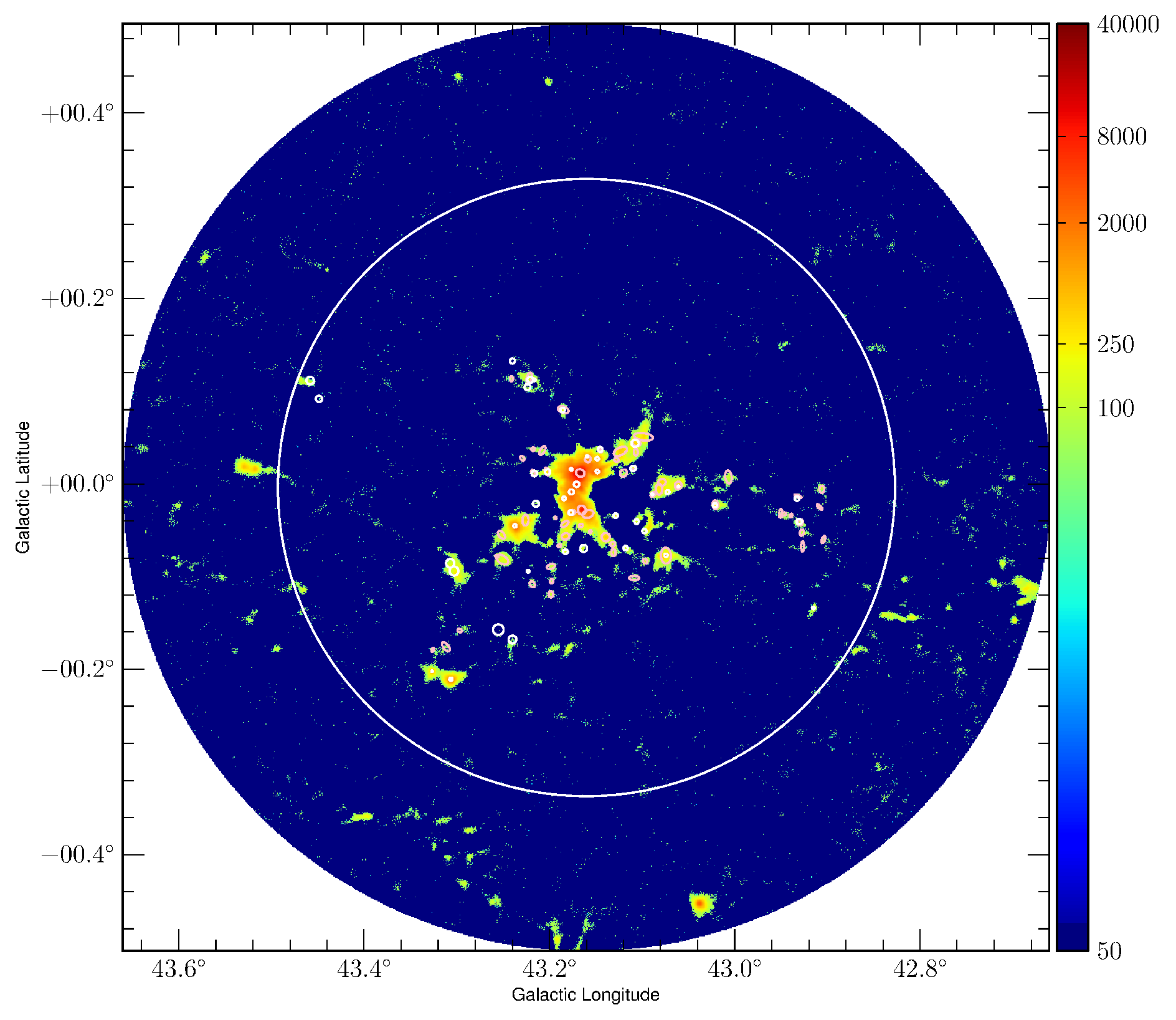}
\caption{The JCMT map of the W49A region. The intensity scale is in units of mJy\,beam$^{-1}$. The white circle indicates a 20-arcmin radius, corresponding to a physical radius of $\sim$\,60\,pc, within which all extracted sources are analysed. The dynamic range is scaled in such a way that only pixels above 3$\upsigma$ are displayed. The pink ellipses represent the JCMT sources assigned to W49A, with the small white circles showing the positions of the Hi-GAL sources.}
\label{W49map}
\end{figure*}

The W49A data were observed in the same method as the JPS, as outlined in \citet{Eden17}, between September 2013 and September 2014 in the weather band with 220-Ghz sky opacities of $\uptau_{220}$\,$\simeq$\,0.08\,--\,0.16, JCMT band-2. The observations consisted of 23 individual $\emph{pong3600}$ observations \citep{Bintley14}, each taking 40-45 minutes and covering a one-degree circular field. The data, reduced with 3-arcsecond pixels using the same procedure described in \citet{Eden17}, have a pixel-to-pixel rms of 17.39\,mJy\,beam$^{-1}$, 4.99\,mJy\,beam$^{-1}$ when smoothed over the beam.  The resulting map is displayed in Fig.~\ref{W49map}.  When utilising the full dymanic range, the data display negative bowling around the bright W49A region, a common feature of the observation and reduction process. For a full explanation, see \citet{Mairs15} and \citet{Eden17}. However, while potentially influencing photometry results in the affected area, this effect does not appear to be a significant factor in the results. The depth of the negative bowling is $\sim$\,10\,$\upsigma$, compared to the $\sim$\,2500\,$\upsigma$ at the brightest point of the data.  This means that few, if any, significant compact sources will have been missed due to this effect.  Additionally, no ATLASGAL compact sources \citep{Urquhart14b} or Hi-GAL band-merged sources \citep{Elia17} are found in the negative regions.
The corresponding W51 map from the JPS is displayed in Appendix~\ref{W51app}.

\subsection{Molecular-line data}

Molecular-line data are available for both regions in the $J = 1-0$ (110.150\,GHz) and $J = 3-2$ (330.450\,GHz) rotational transitions of $^{13}$CO.
The $^{13}$CO $J = 1-0$ data form part of the Galactic Ring Survey (GRS; \citealt{Jackson06}), which mapped the inner Galaxy at Galactic longitudes of $\emph{l}$\,=\,18$\degr$ to 55$\fdg$7 and latitudes of $\emph{b}$\,$\leq$\,1$\degr$, at an angular resolution of 46 arcsecs.

The higher-energy transition of $J = 3-2$ was mapped at an angular resolution of $\sim$14 arcsecs as part of two different projects with the Heterodyne Array Receiver Program (HARP; \citealt{Buckle09}) instrument on the JCMT. The W49A data are part of the $^{13}$CO/C$^{18}$O Heterodyne Inner Milky Way Plane Survey (CHIMPS; \citealt{Rigby16}), while the W51 data are from the targeted survey of the region by \citet{Parsons12}.

\section{Hi-GAL source selection \& source properties}

\subsection{Source selection}

\subsubsection{Hi-GAL sources}

A maximum projected radius of 60\,pc from the main star-forming centre was imposed as the first source-selection criterion in the two regions. The radius of 60\,pc corresponds to approximately double the size of the largest molecular clouds in the GRS catalogue \citep{Roman-Duval10} and to the size of the largest giant molecular clouds in the Galaxy (e.g. W3: \citealt{Polychroni12}), ensuring all material associated with the region is included in this study. Previous studies of W49A were also confined to a radius of 60\,pc \citep{Galvan-Madrid13}. This corresponds to a radius of 20\,arcmin centred on $\emph{l} = 43\fdg$170, $\emph{b}$ = $-$0$\fdg$004 for W49A and a radius of 40\,arcmin from $\emph{l} = 49\fdg$486, $\emph{b}$ = $-$0$\fdg$381 for W51. Next, a source must have a detection in at least 3 of the 4 sub-millimetre wavelengths of the Hi-GAL band-merged catalogue \citep{Elia17}, i.e., 160, 250, 350 and 500\,$\micron$. 148 and 712 candidate Hi-GAL sources were found meeting these criteria for W49A and W51, respectively. 

In order to define association with the target regions by velocity, CO spectra were extracted from the GRS and HARP data cubes at the positions of the above 860 candidate sources. The HARP spectra were inspected first, as the $J = 3-2$ transition traces denser ($\gtrsim$ 10$^{4}$\,cm$^{-3}$) and warmer ($\sim$30\,K) gas than does the $J = 1-0$ transition (10$^{2}$--10$^{3}$\,cm$^{-3}$ and $\sim$10\,K). As the lines of sight towards W49A and W51 contain multiple emission components at different velocities due to foreground and background spiral arms, the $J = 3-2$ transitions are less ambiguous than $J = 1-0$ at identifying the molecular emission associated with a dense, star-forming clump. 176 candidate Hi-GAL sources had emission in the HARP spectra, of which 50 were in W49A and 126 in W51. 

For those candidate sources with multiple emission peaks at different velocities in the spectra, the strongest emission peak was chosen on the assumption that it corresponds to the highest column density (e.g. \citealt{Urquhart07,Eden12,Eden13}). In total, using both HARP and GRS data, 762 sources (121 in W49A and 641 in W51) were assigned velocities.

The velocities obtained from the HARP or GRS spectra were cross-referenced with the \citet{Rathborne09} GRS cloud catalogue, containing the derived distances from \citet{Roman-Duval09}. A full description of the matching method can be found in \citet{Eden12}. These cloud distances were adopted as the distances to the Hi-GAL sources, resulting in assigned distances to 109 and 582 sources in the W49A and W51 target areas, respectively. Of these, 57 and 406 were coincident with the accepted distances of W49A (11.11$^{+ 0.79}_{- 0.69}$\,kpc) and W51 (5.41$^{+ 0.31}_{- 0.28}$\,kpc). The tolerance was taken to be equal to the quoted errors on the cloud distances.

The 71 candidate sources with a $^{13}$CO velocity but without a GRS cloud association were assigned two kinematic distances using the Galactic rotation curve of \citet{Brand93}, due to the distance ambiguity that exists in the Inner Galaxy. Since W51 has a velocity consistent with the rotation tangent point along that particular line of sight ($\sim$ 60\,km\,s$^{-1}$), sources in the W51 field did not require a determination between the two kinematic distances, with all sources at that velocity placed at the tangent distance. 28 out of 59 candidate W51 sources could thus be assigned to the W51 complex on velocity alone. Of the remaining 12 candidate sources with velocities within the W49A field, only two had one kinematic distance consistent with W49A. To determine between the two kinematic distances for these two sources, the HISA method is used \citep[e.g.][]{Anderson09, Roman-Duval09}, making use of \ion{H}{i} from the VLA Galactic Plane Survey \citep{Stil06}. Neither of the two were assigned the far distance, i.e. not determined to be in W49A.

The final source numbers, including all Hi-GAL sources within the selection radii and at the distances of the two regions, are 57 and 434 for W49A and W51, respectively.

\subsubsection{JCMT sources}

The source-extraction process for the new JCMT data makes use of the {\sc FellWalker} (FW; \citealt{Berry15}) algorithm, with the same configuraton parameters used to produce the JPS compact-source catalogue \citep{Eden17}. 173 sources were found in the W49A map, after excluding sources with aspect ratios greater than 5 and SNR less than 5$\upsigma$. A sample of the source catalogue is displayed in Table~\ref{W49sources} (the full list of 173 W49A sources is available as Supporting Information to the online article). Since the observing and source-extraction methods are identical to those used for the JPS, we can estimate the sample completeness limit by scaling the JPS results, with 95 per cent completeness obtained for sources over 5$\upsigma$ \citep{Eden17}, or 86.95\,mJy\,beam$^{-1}$.

The W51 JCMT data, as part of the $\ell$\,=\,50$\degr$ JPS field, have a somewhat higher pixel-to-pixel rms of 25.66\,mJy\,beam$^{-1}$, or 5.98\,mJy\,beam$^{-1}$, when smoothed over the beam \citep{Eden17}. 822 compact sources were found within this JPS field, 384 within 40\,arcmin of the W51 region. Within the same 20-arcmin angular radius, 117 850-$\upmu$m compact sources were found in the W49A map.

The CO spectra at the positions of the JCMT sources were extracted in the same manner as above, with 472 of the 501 candidate JCMT sources assigned velocities, 109 for W49A and 363 for W51. These velocities produced GRS cloud matches, and thus distances, to 61 and 287 sources within the two regions, respectively.  Of the sources without cloud distances, using the methods above, a further 6 of 8 were assigned to W51 and zero of 19 had a far kinematic distance consistent with W49A.

These selection criteria gave 61 and 293 JCMT sources within the W49A and W51 regions, respectively. A summary of the source numbers can be found in Table.~\ref{sourcenumbers}. The big difference in the source numbers, in both the Hi-GAL and JCMT samples, is probably due to source blending at the greater distance of W49A.  This issue is addressed below (section 4.2). The source IDs of the Hi-GAL and JPS sources used are listed in Appendix~\ref{sourceIDs}.

\begin{table}
\begin{center}
\caption{Summary of the source numbers found in each survey for W49A and W51.}
\label{sourcenumbers}
\begin{tabular}{lcc}\hline
Region & Hi-GAL & JCMT \\
& Sources & Sources \\
\hline
W49A & 57 & 61  \\
W51 & 434 & 293 \\
\hline
\end{tabular}
\end{center}
\end{table}

\begin{table*}
\begin{center}
\caption{The W49A JCMT catalogue. The columns are as follows: (1) W49A catalogue source name; (2) and (3) Galactic coordinates of the position at which the peak flux is found within the W49A source; (4) and (5) the central point in Galactic coordinates; (6--8) semi-major, semi-minor and position angle, measured anticlockwise from the Galactic north, of the elliptical fit to the shape of the source; (9) effective radius of source, calculated by $\sqrt{(A/\pi)}$, where A is the area of the source above the FW detection threshold; (10--11) peak flux density, in units of Jy\,beam$^{-1}$, and measurement error; (12--13) integrated flux, in units of Jy, and measurement error; (14) signal-to-noise ratio (SR) of the source and (15) whether the source is associated with the W49A star-forming region, determined by heliocentric distance.}
\label{W49sources}
\begin{tabular}{lccccccccccccccc}\hline
Source & $\ell_{\rmn{peak}}$ & $\emph{b}_{\rmn{peak}}$ & $\ell_{\rmn{cen}}$ & $\emph{b}_{\rmn{cen}}$ & $\upsigma_{\rmn{maj}}$ & $\upsigma_{\rmn{min}}$ & PA & $R_{\rmn{eff}}$ & $S_{\rmn{peak}}$ & $\Delta$$S_{\rmn{peak}}$ & $S_{\rmn{int}}$ & $\Delta$$S_{\rmn{int}}$ & SNR & W49A\\
ID & ($^{\circ}$) & ($^{\circ}$) & ($^{\circ}$) & ($^{\circ}$) & ($\prime\prime$) & ($\prime\prime$) & ($^{\circ}$) & ($\prime\prime$) & (Jy\,beam$^{-1}$) & (Jy\,beam$^{-1}$) & (Jy) & (Jy) &  & Source\\
(1) & (2) & (3) & (4) & (5) & (6) & (7) & (8) & (9) & (10) & (11) & (12) & (13) & (14) & (15) \\
\hline
W49\_021	&	42.871	&	-0.182	&	42.866	&	-0.179	&	18	&	9	&	190	&	26	&	0.145	&	0.027	&	0.577	&	0.029	&	8.31	&	n	\\
W49\_022	&	42.884	&	-0.030	&	42.882	&	-0.029	&	10	&	7	&	206	&	17	&	0.090	&	0.017	&	0.175	&	0.009	&	5.17	&	n	\\
W49\_023	&	42.888	&	-0.082	&	42.887	&	-0.083	&	10	&	4	&	137	&	13	&	0.088	&	0.016	&	0.089	&	0.005	&	5.07	&	n	\\
W49\_024	&	42.888	&	-0.193	&	42.886	&	-0.190	&	8	&	7	&	212	&	15	&	0.103	&	0.019	&	0.157	&	0.008	&	5.95	&	n	\\
W49\_025	&	42.889	&	-0.197	&	42.888	&	-0.197	&	15	&	6	&	137	&	17	&	0.108	&	0.020	&	0.195	&	0.010	&	6.18	&	n	\\
W49\_026	&	42.904	&	-0.060	&	42.902	&	-0.061	&	15	&	6	&	160	&	19	&	0.098	&	0.018	&	0.241	&	0.012	&	5.64	&	y	\\
W49\_027	&	42.906	&	-0.006	&	42.908	&	-0.005	&	14	&	9	&	177	&	25	&	0.125	&	0.024	&	0.453	&	0.023	&	7.19	&	y	\\
W49\_028	&	42.908	&	-0.025	&	42.907	&	-0.024	&	14	&	5	&	225	&	17	&	0.097	&	0.018	&	0.192	&	0.010	&	5.59	&	y	\\
W49\_029	&	42.915	&	-0.134	&	42.915	&	-0.135	&	13	&	7	&	241	&	22	&	0.200	&	0.037	&	0.498	&	0.025	&	11.52	&	n	\\
W49\_030	&	42.922	&	-0.142	&	42.920	&	-0.143	&	6	&	6	&	230	&	12	&	0.093	&	0.018	&	0.098	&	0.005	&	5.32	&	n	\\
W49\_031	&	42.927	&	-0.067	&	42.932	&	-0.068	&	20	&	6	&	179	&	21	&	0.090	&	0.017	&	0.246	&	0.012	&	5.18	&	y	\\
W49\_032	&	42.928	&	-0.052	&	42.924	&	-0.053	&	12	&	8	&	169	&	20	&	0.126	&	0.024	&	0.329	&	0.016	&	7.27	&	y	\\
W49\_033	&	42.929	&	-0.042	&	42.930	&	-0.042	&	12	&	9	&	158	&	24	&	0.196	&	0.037	&	0.601	&	0.030	&	11.28	&	y	\\
W49\_034	&	42.932	&	-0.013	&	42.930	&	-0.012	&	14	&	5	&	247	&	18	&	0.092	&	0.017	&	0.178	&	0.009	&	5.29	&	y	\\
W49\_035	&	42.939	&	-0.034	&	42.940	&	-0.034	&	7	&	5	&	151	&	13	&	0.113	&	0.021	&	0.181	&	0.009	&	6.48	&	y	\\
W49\_036	&	42.945	&	-0.314	&	42.943	&	-0.313	&	8	&	4	&	100	&	12	&	0.088	&	0.017	&	0.092	&	0.005	&	5.08	&	n	\\
W49\_037	&	42.946	&	0.149	&	42.946	&	0.151	&	9	&	6	&	188	&	16	&	0.096	&	0.019	&	0.187	&	0.009	&	5.50	&	n	\\
W49\_038	&	42.947	&	-0.032	&	42.946	&	-0.034	&	8	&	6	&	101	&	15	&	0.109	&	0.021	&	0.203	&	0.010	&	6.26	&	y	\\
W49\_039	&	42.950	&	-0.032	&	42.956	&	-0.030	&	16	&	6	&	181	&	20	&	0.098	&	0.019	&	0.249	&	0.012	&	5.61	&	y	\\
W49\_040	&	42.951	&	0.147	&	42.951	&	0.147	&	10	&	5	&	126	&	15	&	0.092	&	0.018	&	0.136	&	0.007	&	5.28	&	n	\\
\hline
\multicolumn{15}{l}{$\emph{Note:}$ Only a small portion of the data is provided here, with the full list of 173 W49A sources available as Supporting Information to the online article. }\\
\end{tabular}
\end{center}
\end{table*}

\subsection{Luminosity determination}

The luminosities of the Hi-GAL sources are given in the Hi-GAL compact-source catalogue \citep{Elia17} and were calculated by fitting a modified blackbody to the spectral energy distribution (SED) of each source above $21\,\umu$m, using the fitting strategy as described in \citet{Giannini12}. The SED fitting and consequent luminosity calculations are fully explained in \citet{Elia17}, with a brief description below.

The justification for the use of a modified blackbody as opposed to an SED template is described in \citet{Elia16}. The modified blackbody expressions, and adopted constants, are as described by \citet{Elia13}.  
Note that, in order to account for the different angular resolutions in each band, the fluxes at 350 and 500\,$\micron$ are scaled by the ratio of the beam-deconvolved source sizes in each band to that at 250\,$\micron$ (e.g., \citealp{Giannini12}; cf \citealp{NguyenLuong11})

The value of the dust opacity exponent, $\beta$, is kept constant at 2.0 in the fit, as recommended by \citet{Sadavoy13} in the $\emph{Herschel}$ Gould Belt Survey and as adopted in the HOBYS survey \citep{Giannini12}. The integrated flux is then converted to luminosity, $L$, and temperature, $T_{\rmn{d}}$, which are free parameters.  The luminosities of the sources also include shorter, and longer, wavelength components detected with various other surveys, allowing for these values to approximate bolometric luminosities. Shorter-wavelength surveys used include MIPSGAL \citep{Gutermuth15}, MSX \citep*{Egan03}, and WISE \citep{Wright10}, whilst longer wavelengths made use of the GaussClumps ATLASGAL catalogue \citep{Csengeri14} and the version 2 catalogue of the BGPS \citep{Ginsburg13}. The use of the \citet{Csengeri14} ATLASGAL catalogue emphasises the compact nature of the Hi-GAL sources as these ATLASGAL sources are more compact than those of \citet{Contreras13} and \citet{Urquhart14b}. The completeness limits of the luminosities correspond to 200\,L$_{\sun}$ and 100\,L$_{\sun}$ for W49A and W51, respectively.

\begin{figure}
\includegraphics[scale=0.50]{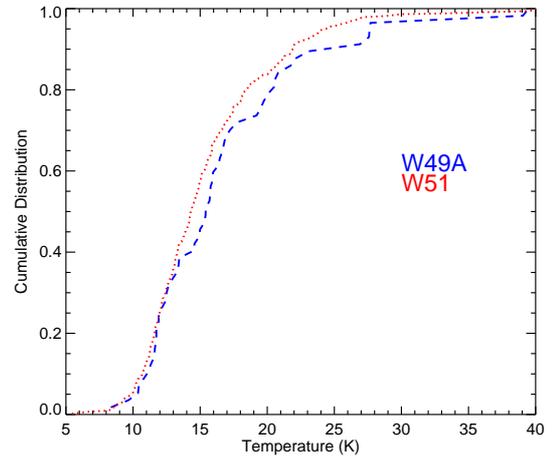}
\caption{The cumulative distributions of the SED-derived temperatures for Hi-GAL sources in the W49A and W51 regions, represented by the blue dashed and red dotted lines, respectively.}
\label{temperatures}
\end{figure}

The cumulative distribution of the fitted temperatures in the two regions is shown in Fig.~\ref{temperatures}. The mean temperatures are 16.8\,$\pm$\,0.8\,K and 15.4\,$\pm$\,0.2\,K with median temperatures of 15.4\,$\pm$\,3.5\,K and 14.3\,$\pm$\,2.6\,K for W49A and W51, respectively.  A Kolomogorov--Smirnov (K--S) test was applied to the $T_{\rmn{d}}$ distributions of the two sub-samples giving a 22 per cent probability that the differences arise from random sampling fluctuations, so it can be assumed that these subsets are similarly evolved.

\subsubsection{Method Dependency of Luminosities}

The luminosities quoted in this study are those given in \citet{Elia17} and are obtained using the method outlined in that paper. The total luminosities contained within all compact clumps are found to be 1.03\,$\times$\,10$^{6}$\,L$_{\sun}$ and 4.67\,$\times$\,10$^{5}$\,L$_{\sun}$ for W49A and W51, respectively. These values are an order of magnitude smaller than those found in other studies.  For example, \citet{Urquhart17} find integrated compact-source luminosities of 1.52\,$\times$\,10$^{7}$\,L$_{\sun}$ and 1.11\,$\times$\,10$^{7}$\,L$_{\sun}$, respectively. As described above, the Hi-GAL luminosities of \citet{Elia17} use fluxes scaled to the source size at 250\,$\upmu$m, which will remove flux at longer wavelengths and for larger sources.  The corresponding fluxes in the \citet{Urquhart17} 
study, extracted from the public Hi-GAL image data, use a 3-$\upsigma$ aperture radius, which corresponds to a minimum aperture size of 55.1 arcsec \citep{Konig17}.

Total integrated luminosities obtained from the image data, rather than by adding the compact sources, including all emission in all wavebands within 60\,pc radii, but otherwise calculated as in \citet{Elia17}, are 8.82\,$\times$\,10$^{6}$\,L$_{\sun}$ and 1.45\,$\times$\,10$^{6}$\,L$_{\sun}$ for W49A and W51, respectively. 
These are consistent with the literature values quoted in the introduction \citep{Harvey77,Kang10}. This consistency implies that the low values of $L$ are the result of the aperture-photometry methodology of \cite{Elia17}.  However, this emphasises that the derived luminosity distributions are strictly relevant to compact sources, at the position where the YSO is most likely to form within a clump, and tend to exclude extended emission.

\subsection{Mass determination}

The masses of the JCMT detected sources were calculated using the following:

\begin{equation}
M = \frac{S_{\nu}D^{2}}{\kappa_{\nu}B_{\nu}(T_{\rmn{d}})}
\end{equation}

\noindent where $S_{\nu}$ is the integrated flux density, $D$ is the distance to the source, $\kappa_{\nu}$ is the mass absorption coefficient, taken to be 0.001\,m$^{2}$\,kg$^{-1}$ at a wavelength of 850\,$\upmu$m \citep{Mitchell01} and $B_{\nu}(T_{\rmn{d}})$ is the Planck function evaluated at a dust temperature, $T_{\rmn{d}}$. Taking the distances to W49A and W51 as 11.11\,kpc and 5.41\,kpc, respectively \citep{Sato10,Zhang13}, and the dust temperatures as the median values from above (15.37\,K and 14.28\,K, respectively), the equation becomes $M/M_{\sun} = 2066\,S_{\nu}/\rmn{Jy}$ and $M/M_{\sun} = 490\,S_{\nu}/\rmn{Jy}$ for the two regions, respectively. The masses are calculated from the JCMT data to maintain some independence between the determination of $M$ and $L$. The median temperatures are used in the instances where there are not positional matches, within the $\emph{Herschel}$ beam, with a Hi-GAL source. Where there is a match, the SED-derived temperature is used. The SED-derived temperatures are used in 37 and 148 cases for W49A and W51, respectively.

We can compare the masses derived from the JCMT single fluxes to those of the ATLASGAL survey \citep{Urquhart17}, which were derived from SED fits. We find for W49A, 2.54\,$\times$\,10$^{5}$\,M$_{\odot}$ and 2.26,$\times$\,10$^{5}$\,M$_{\odot}$ for the SCUBA-2 masses and ATLASGAL masses, respectively, and 2.49\,$\times$\,10$^{5}$\,M$_{\odot}$ and 2.12\,$\times$\,10$^{5}$\,M$_{\odot}$, respectively. This allows us to confidently say our masses are a good estimate of the sub-mm mass in the two regions, whilst maintaining the independence of $M$ and $L$.

\section{Results}

\subsection{Clump mass and luminosity distributions}

\begin{figure}
\includegraphics[scale=0.50]{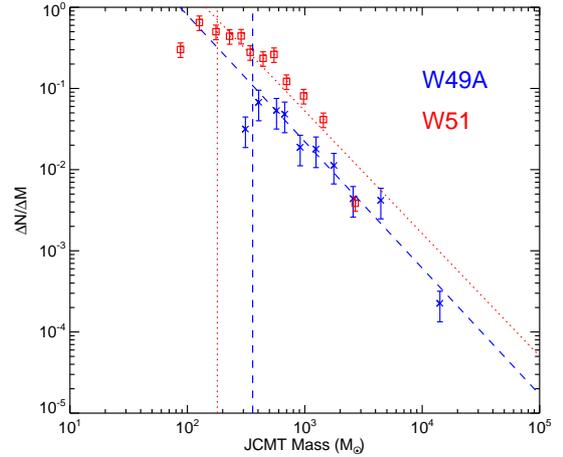}
\caption{The clump mass distributions of W49A and W51 plotted with blue crosses and red squares, respectively, with the masses derived from the JCMT sub-millimetre fluxes and Equation 1. The least-squares fit for each CMD is overlaid with a dashed and dotted line, respectively. The vertical lines in blue (dashed) and red (dotted) are the sample completeness limits for W49A and W51, respectively.}
\label{massfunctions}
\end{figure}

\begin{figure}
\includegraphics[scale=0.50]{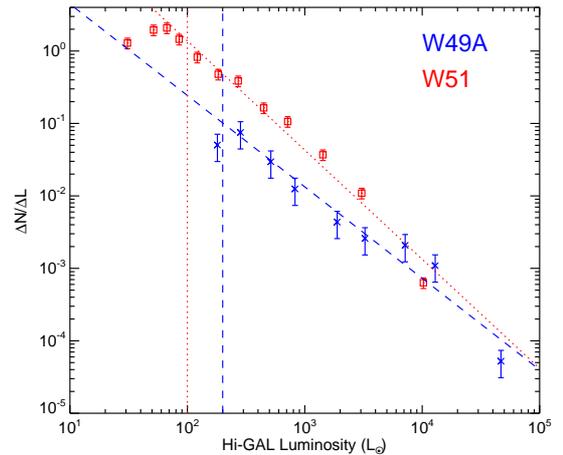}
\caption{The luminosity distributions of W49A and W51 plotted with blue crosses and red squares, respectively, with the luminosities derived from the Hi-GAL SED fits. The least-squares fit for each LD is overlaid with a dashed and dotted line, respectively. The vertical lines in blue (dashed) and red (dotted) are the sample completeness limits for W49A and W51, respectively.}
\label{lumfunctions}
\end{figure}

Using the luminosities and masses derived in Sections 3.2 and 3.3, clump mass distributions (CMDs) and luminosity distributions (LDs) are plotted, which are presented in Fig.~\ref{massfunctions} and Fig~\ref{lumfunctions}, respectively.

The plotted quantities, $\Delta\emph{N/}\Delta\emph{M}$ and $\Delta\emph{N/}\Delta\emph{L}$, are the number of sources per mass or luminosity bin width, with the mass and luminosity coordinate represented by the median value in each bin. This method was used to plot LDs in \citet{Eden15}. A fixed number of sources per bin was used, as opposed to fixed bin widths, in order to equalise weights determined from Poisson errors \citep{MaizApellaniz05}.

By assuming a power-law slope of the form $\Delta\emph{N/}\Delta\emph{M}$ $\propto$ $M^{\alpha}$ and $\Delta\emph{N/}\Delta\emph{L}$ $\propto$ $L^{\gamma}$, least-squares fit to both CMDs and LDs can be calculated. Indices of $\alpha = -1.55 \pm 0.11$ and $\alpha = -1.51 \pm 0.06$ are calculated for the CMDs for W49A and W51, respectively, and $\gamma = -1.26 \pm 0.05$ and $\gamma = -1.51 \pm 0.03$ for the LDs for W49A and W51, respectively. The fits are performed on all bins above the completeness limit calculated from the 5-$\upsigma$ rms noise in the JPS data (CMDs) and the 95 per cent detection limit in the Hi-GAL data (LDs; \citealt{Molinari16a}). These limits are taken to be 360\,M$_{\sun}$ and 200\,L$_{\sun}$ for the W49A mass and luminosity distributions, respectively, and 180\,M$_{\sun}$ and 100\,L$_{\sun}$ for the W51 data.

The fitted index values for the two CMDs are consistent with each other but those of the LDs are statistically different at the 5-$\upsigma$ level, with the W49A luminosity distribution being more top-heavy (flatter). The LD of W51 is consistent with those found for YSOs in Galactic-wide samples ($\alpha = -1.50 \pm 0.02$; \citealt{Mottram11,Urquhart14a}, $\alpha = -1.57 \pm 0.07$; \citealt{Eden15}), that in nearby clouds ($\alpha = -1.41 \pm 0.03$; \citealt{Kryukova12}), and the Cygnus-X and W43 star-forming regions ($\alpha = -1.63 \pm 0.03$; \citealt{Kryukova14}, $\alpha = -1.55 \pm 0.05$; \citealt{Eden15}). It is, however, worth noting that the final point of the W51 LD is constraining the fit. When a fit is performed without that point, it is significantly shallower and consistent with W49A and so both are flatter, in this case, than the Galactic average.  The CMDs found for each region are consistent with the Galactic mean \citep{Beuret16,Elia17}.

Monte Carlo simulations of the slopes of the LDs provide an estimate of how the observational errors of the individual luminosities propagate. We combined the uncertainties on the individual luminosities, taken to be 30 per cent (D.\ Elia, private communication), as well as any uncertainties on the association with the two regions. The luminosity of each source in the LD was then sampled from within these error bars, and a new LD produced, with a calculated fit. This was repeated 1000 times. We find that this analysis gives the errors on the LDs as 0.049\,$\pm$\,0.003 and 0.032$\pm$\,0.001 for W49A and W51, respectively. Therefore we conclude that these observational errors are not altering the derived slopes significantly.

\subsection{Mass-luminosity relationship}

\begin{figure}
\includegraphics[scale=0.50]{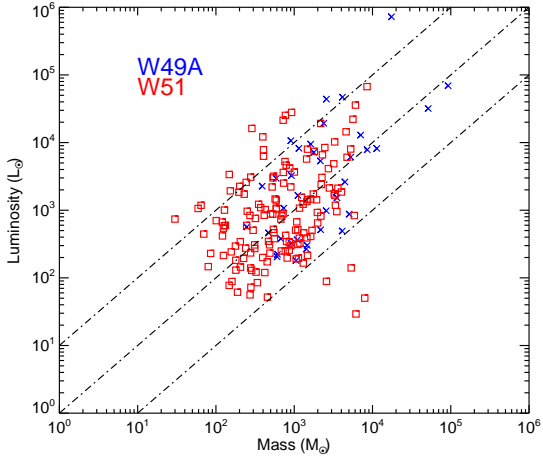}
\caption{Mass-luminosity relationship for W49A and W51, represented by the blue crosses and red squares, respectively, with fits to those data indicated by the dashed and dotted lines, respectively. The lower, middle and upper black dot-dash lines represent the $L/M_{\rmn{clump}}$ = 0.1, 1 and 10\,L$_{\sun}$/M$_{\sun}$, respectively.}
\label{M-L}
\end{figure}

The JCMT clumps were positionally matched to Hi-GAL YSOs with a tolerance of 40$\arcsec$, approximately the $\emph{Herschel}$ beam FWHM at 500\,$\upmu$m. This resulted in 37 JPS clumps (matched with 44 Hi-GAL sources) in W49A and 148 JPS clumps (matched with 267 Hi-GAL sources) in W51. The relationship between the mass of a clump and luminosity of the associated Hi-GAL infrared source is shown in Fig.~\ref{M-L}, for both regions. There is very little correlation found in both samples, with Spearman-rank tests giving correlation coefficients of 0.58 and 0.34 for W49A and W51, respectively, with associated $p$-values of 0.24 and 0.21. The lack of correlation is possibly due to the narrow range of $L$ and $M$ found within individual regions, compared to the well constrained correlations found across many orders of magnitude of $L$ and $M$ in much larger samples \citep[e.g][]{Urquhart14}.

\subsection{Distance effects}

One potential source of bias affecting the LDs of the two regions is that W49A is at approximately double the distance of W51 (11.11\,kpc compared to 5.41\,kpc). The CMDs are not subject to these biases as studies have found that the slopes of CMDs do not change across different distance ranges, both heliocentric and Galactocentric \citep[e.g][]{Eden12,Elia17}. One potential effect is seen in observations and simulations \citep{Moore07,Reid10} in which the clustering scale of the sources and the angular resolution of the survey combine to bump lower-mass clumps into the higher bins. The CMDs and LDs show evidence of this but, as seen in \citet{Reid10}, the slope before and after these ``bumps'' in the distributions is the same as the high-mass clumps are rare and do not generally get merged with each other, so the high-end slope would not be affected, except in extreme cases.

To mitigate the effects of distance, we use the method outlined in \citet{Baldeschi17} to simulate placing the W51 region at the same distance as W49A. This method rescales and rebins the map according to the ratio of the distances. The rescaled W51 map is then convolved with the point-spread function of the instrument, again scaled by the relative distance. After which, noise is added to the map to replicate the noise that was reduced in the smoothing process. The ``moved'' map was then subject to the same CuTEx source extraction and SED fitting as the original Hi-GAL maps. In the rescaled map, 134 sources were extracted. However, as the real velocities are no longer relevant, all sources within the angular radius of 20\,arcmins were assigned to the W51 star-forming region. The number of sources is similar to the number of sources found in W49A, indicating the potential source blending in action in W49A. The luminosities are shifted by an order of magnitude compared to the original W51 map, with the highest luminosity sources consistent with W49A. 

\begin{figure}
\includegraphics[scale=0.50]{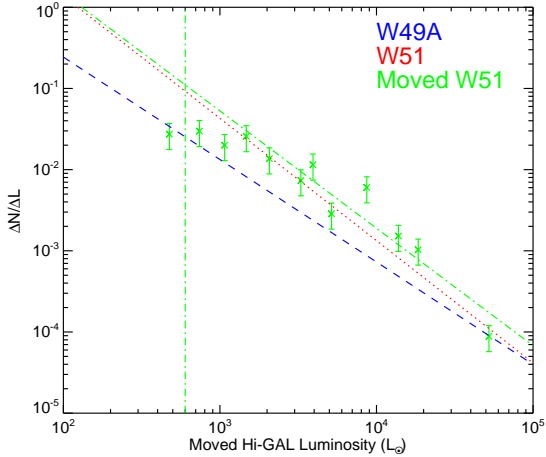}
\caption{The luminosity distribution of W51 after resampling and smoothing the data and re-extracting the sources, using the technique of \citet{Baldeschi17}, to simulate it being at the same distance as W49A.  The binned data are represented by the green crosses, with the dot-dash line showing the linear best fit. The gradients of the luminosity distributions in W49A and in the original W51 data are displayed with the blue dashed and red dotted lines, respectively. The vertical, green dot-dash line represents the completeness limit.}
\label{movedLD}
\end{figure}

These sources were then used to calculate the LD of the moved W51, and the index was found to be $\gamma = -1.45 \pm 0.07$ above 600\,M$_{\sun}$, with the LD shown in Fig.~\ref{movedLD}. This value is consistent with that of the original W51 LD and is still significantly steeper than that of the W49A molecular cloud.

\section{Discussion}

\subsection{A comparison of W49A and W51}

A number of the quantities that are commonly used to compare the star-forming content of different regions have been calculated for W49A and W51 and are displayed in Table~\ref{quantities}. These parameters are the indices of power-law fits to the CMDs and LDs, as derived in Section 4.1; the ratio of infrared luminosity to the clump; the clump formation efficiency (CFE), the percentage of molecular gas that was converted to dense, star-forming material; the number of infrared sources per unit cloud mass; and star-forming fraction (SFF), the number of Hi-GAL sources with an associated 70-$\upmu$m source \citep{Ragan16}. The molecular gas masses are taken from \citet{Galvan-Madrid13} and \citet{Roman-Duval10} for W49A and W51, respectively. Included in the cloud mass for W51 are those clouds associated with the sources as well as clouds at the distances W51 but within the on-sky selection radii. The additional clouds at the distance of W49A were accounted for by \citet{Galvan-Madrid13}, who derived the molecular mass within a radius of 60\,pc.

The CFE, as defined in \citet{Eden12,Eden13}, takes a snapshot of the current star formation and any variation in this quantity implies either an altered timescale for clump formation, or a change in the clump-formation rate. As the clump formation stage is short \citep[e.g.][]{Mottram11}, we assume that any change is due to an altered rate.

These quantities cover the scale of the whole cloud (CFE, infrared sources per unit cloud mass) to the scale of individual clumps (SFF, ratio of infrared luminosity to clump mass). By covering these scales, we can identify if changes in quantities are associated with a specific stage of star formation.

\begin{table*}
\begin{center}
\caption{Summary of the quantities calculated for the W49A and W51 star-forming regions, with the W51 moved results included as well as Galactic averages, alongside the relevant reference.}
\label{quantities}
\begin{tabular}{lccccc} \hline
Parameter & W49A & W51 & Moved W51 & Galactic Avg. & Reference \\
\hline
Index of CMD & -1.56\,$\pm$\,0.11 & -1.51\,$\pm$\,0.06 & --- & -1.57\,$\pm$\,0.07 & \citealt{Beuret16} \\
Index of LD & -1.26\,$\pm$\,0.05 & -1.51\,$\pm$\,0.03 & -1.45\,$\pm$\,0.07 & -1.50\,$\pm$\,0.02 & \citealt{Eden15} \\
$L_{\rmn{IR}}/M_{\rmn{clump}}$ (L$_{\sun}$\,M$_{\sun}^{-1}$) & 3.12\,$\pm$\,0.59 & 3.52\,$\pm$\,0.34 & --- & 1.39\,$\pm$\,0.09 & \citealt{Eden15} \\
Mean $L_{\rmn{IR}}/M_{\rmn{clump}}$ (L$_{\sun}$\,M$_{\sun}^{-1}$) & 3.80\,$\pm$\,1.22 & 4.05\,$\pm$\,0.62 & --- & 5.24\,$\pm$\,0.70  & \citealt{Eden15} \\
Median $L_{\rmn{IR}}/M_{\rmn{clump}}$ (L$_{\sun}$\,M$_{\sun}^{-1}$) & 0.91\,$\pm$\,0.71 & 1.21\,$\pm$\,0.93 & --- & 1.72\,$\pm$\,1.14  & \citealt{Eden15} \\
$M_{\rmn{clump}}/M_{\rmn{cloud}}$ (per cent) & 62.3\,$\pm$\,13.7 & 39.9\,$\pm$\,6.0 & --- & 11.0\,$\pm$\,6.0 & \citealt{Battisti14} \\
YSOs per Cloud Mass ($\times$10$^{-4}$\,M$_{\sun}^{-1}$) & 0.90\,$\pm$\,0.17 & 6.94\,$\pm$\,1.03 & 2.14\,$\pm$\,0.35 & 0.05\,$\pm$\,0.01 & \citealt{Moore12} \\
SFF & 0.29\,$\pm$\,0.05 & 0.30\,$\pm$\,0.02 & --- & 0.25 & \citealt{Ragan16} \\
\hline
\end{tabular}
\end{center}
\end{table*}

Some quantities associated with the actively star-forming evolutionary stage show a variation between the two regions.  The index of the power-law fitted to the luminosity distribution was found to be $\gamma = -1.26 \pm 0.05$ in W49A, compared to $\gamma = -1.51 \pm 0.03$ for W51. The LD of W49A is significantly flatter than those found in other star-forming regions and across the Galaxy by \citet{Eden15} and in the RMS survey \citep{Lumsden13,Urquhart14a}, while that of W51 is consistent with those large-scale samples. \citet{Moore12} postulated that a flatter LD in W49A might contribute to measured increases in large-scale $L/M$ in the Perseus spiral arm. They also found lesser but similar increases in $L/M$ associated with the Sagittarius spiral arm due to the inclusion of the W51 region. However, it was suggested that the latter was more likely to be due to an increase in the number of YSOs per unit gas mass. In the present data, we find values of the latter parameter to be $(0.90 \pm 0.17) \times 10^{-4}$ and $(6.94 \pm 1.03) \times 10^{-4}$\,M$_{\sun}^{-1}$ for W49A and W51, respectively. Distance and resolution do affect the latter, as the value for the moved W51 map was found to be $(2.14 \pm 0.35) \times 10^{-4}$\,M$_{\sun}^{-1}$, almost enough to account for the difference between the two regions. The corresponding SFF values are consistent with each other, as well as with the global mean of the inner Galactic Plane \citep{Ragan16}.

The consistency of the W51 LD with the average found in Galaxy-wide samples of high-mass star-forming regions (e.g., \citealt{Mottram11}) suggests that W51 is normal in this regard and its invariance with simulated distance indicates that the flatter slope seen in W49A is not the result of distance-related resolution effects.  W49A therefore appears to be unusual, and may contain a shallow cluster mass function or top-heavy underlying stellar IMF  The high-mass stellar IMF has been found to be invariant within the measurement uncertainties across multiple environments, from the Milky Way to the extremity of starburst-Galaxies \citep{Bastian10}, so any evidence of variations is significant for star-formation theories. 

Quantities associated with the clump formation stage are consistent between the two star-forming complexes. The CMDs have power-law indices that are statistically indistinguishable from each other, which is consistent with the result of \citet{Eden12} who found no variation in CMDs across different Galactic environments, including the W43 star-forming region, and that of \citet{Beuret16} who measured consistent CMDs between clustered and non-clustered clumps. The CFE does not vary between the two regions, again consistent with \citet{Eden12} and \citet{Eden13}. They found this ratio to be constant on average across kiloparsec scales but that large local variations occur, with the distribution of the CFE of individual molecular clouds being consistent with being log-normal. The implication of this is that the most extreme regions are not necessarily abnormal but simply lie in the wings of a  distribution resulting from multiple, multiplicative random processes. The CFEs found for the two regions, $\sim$ 62 and 40 per cent, respectively, are at the high end of these distributions, but comparable with the peak value found in W43 (58\,$\pm$\,13 per cent; \citealt{Eden12}).

\begin{figure}
\includegraphics[scale=0.5]{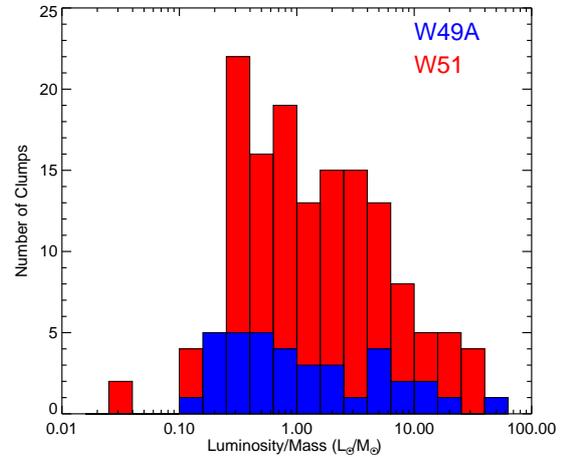}
\caption{Histogram of the ratio of luminosity to mass for individual clumps in the star-forming regions W49A and W51, represented by red and blue bars, respectively.}
\label{L/M}
\end{figure}

The mean values of the star-formation-efficiency analogue, $L/M$, using the clump mass, are also consistent between the two regions. Values for $L_{\rmn{IR}}/M_{\rmn{clump}}$ are found to be $3.12 \pm 0.59$\,L$_{\sun}$\,M$_{\sun}^{-1}$ and $3.52 \pm 0.34$\,L$_{\sun}$\,M$_{\sun}^{-1}$, for W49A and W51, respectively. The values of $L_{\rmn{IR}}/M_{\rmn{clump}}$ compare to the ratio of $1.65 \pm 0.07$\,L$_{\sun}$\,M$_{\sun}^{-1}$ found for W43 \citep{Eden15}. The distribution of $L/M$ values in the two regions is not statistically distiguishable from a log-normal distribution, with Anderson-Darling giving probabilities of 0.15 and 0.15 for the W49A and W51 regions, respectively, with the probablities of the Shapiro-Wilk test found to be 0.11 and 0.10, respectively. This distribution is consistent with those found in a wider sample by \citet{Eden15}, with a log-normal fit giving means of 0.57 and 1.19, with standard deviations of 0.88\,dex and 0.66\,dex for W49A and W51, respectively. However, there is marginal evidence that the inner regions of W49A are different to those on the outer edge in this parameter. Splitting the sample by the median radius from the centre, the distributions of $L/M$ differ at the 2.5-$\upsigma$ level. There is also a hint of bimodality in the W49A sample (Fig.~\ref{L/M}), although the significance is low, with Hartigan's dip test giving a probability of 0.04 that the observed distribution arises at random.

The $L/M$ parameter is both a metric of evolutionary state and an SFE analogue. If the IMF is fully sampled, and the timescale of the selected evolutionary stage (i.e. IR-bright) is short enough to be a snapshot of current star formation, then the $L/M$ of a sample should be proportional to the SFE. However, for a single source, it may be useful to trace the evolution.

The $L-M$ relationship can also be used as an evolutionary indicator of the YSO, and the stage it is in, as it evolves towards the main sequence (\citealt{Molinari08}; \citealt{Giannetti13}). A full description of the evolutionary tracks that a YSO can take can be found in \citet{Molinari08}. It is clear, however, that the two star-forming regions are indistinguishable using this measure, and it is known that radio-faint massive YSOs and \ion{H}{ii} regions occupy the same position in the $M-L$ plane \citep{Urquhart14}. There is evidence though that the star formation in W49 is at a younger stage compared to W51, as well as W43 \citep{Saral15,Saral17}. This is in contrast to the wider Galactic environments in which the two regions are located. \citet{Eden15} found that star formation has distinct time gradients across different Galactic spiral arms, with the star formation in the Perseus arm found to be at a more evolved stage than the other star-forming regions. However, as the clump-formation stage is short, with the onset of star formation almost instantaneous, any differences found at the clump level should indicate a difference in the star formation.

The distributions of the value of $L/M$ in individual clumps (Fig.~\ref{L/M}) are statistically indistinguishable.  The median $L/M$ values are $0.91 \pm 0.71$\,L$_{\sun}$\,M$_{\sun}^{-1}$ and $1.21 \pm 0.93$\,L$_{\sun}$\,M$_{\sun}^{-1}$ for W49A and W51, respectively. The mean values also do not differ significantly, being $3.80 \pm 1.22$\,L$_{\sun}$\,M$_{\sun}^{-1}$ for W49A and $4.05 \pm 0.62$\,L$_{\sun}$\,M$_{\sun}^{-1}$ for W51 (Table \ref{quantities}). A K--S test of the two samples gives a probability of 86 per cent that they are drawn from the same population. These values are consistent with a much wider Galactic sample \citep{Eden15}, which were calculated in a similar way to this study.

\begin{figure}
\includegraphics[scale=0.5]{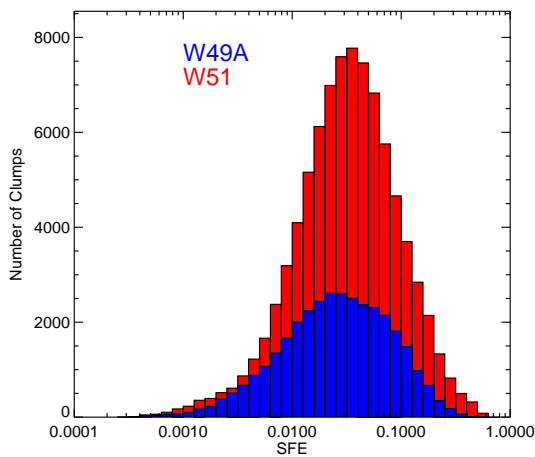}
\caption{Simulated SFEs for matched sources in W49A and W51 with masses above 500\,M$_{\sun}$, represented by red and blue bars, respectively. Each source is run 1000 times.}
\label{truecomp}
\end{figure}

If $L/M$ is the same but the LD is flatter, as is the case in W49A, one would predict that the underlying SFE, i.e., the ratio of stellar mass to either clump or cloud mass, is lower. The probability distribution of the ``true'' SFE of the two regions can be estimated by simulating the populating of an IMF using the Monte-Carlo model of \citet{Urquhart13}. By assuming a standard IMF \citep{Kroupa01}, and halting the random sampling once either the mass of the clump is exceeded, or the observed $L/M$ is, a value for the SFE consistent with these two constraints is recorded.  This is repeated 1000 times for each clump considered in Fig.~\ref{L/M} with a mass of above 500\,M$_{\sun}$, leaving 34 and 86 sources in W49A and W51, respectively. The results of these Monte Carlo simulations are probability distributions for the SFE within each clump which, when added together, provide a probability distribution for the clump SFE in the whole region.  These distributions are presented as histograms in Fig.~\ref{truecomp}.  Gaussian fits to these distributions find that the peak probability lies at SFEs of 2.7 per cent and 3.4 per cent for W49A and W51, respectively. However, these peaks correspond to $\log({\rm SFE}) = -1.56 \pm 0.10$ and $-1.47 \pm 0.04$ for W49A and W51, respectively, and are indistinguishable.

\subsection{$L/M$ and SFE as a function of radius}

\begin{figure}
\includegraphics[scale=0.50]{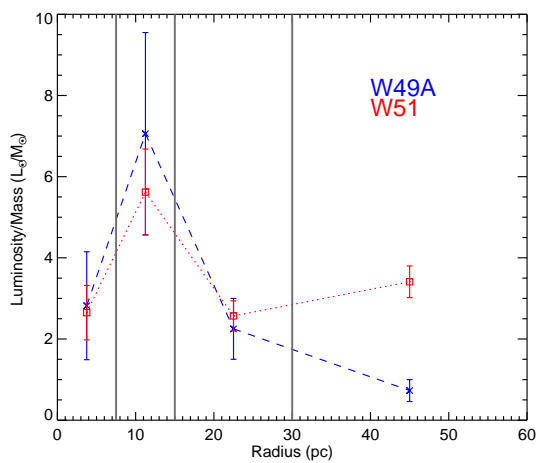}
\caption{Total $L/M$ ratios as a function of radius from the central point of the W49A and W51 regions, represented by the blue crosses and red squares, respectively. The grey vertical lines indicate the boundaries of the radius bins used to calculate the ratios, with the $x$-axis positions representing the centres of each bin.}
\label{rings}
\end{figure}

Radii equivalent to physical sizes of 7.5, 15, 30 and 60\,pc were placed around the central points of W49A and W51, with the total luminosity and clump mass contained within sources within each of these rings summed, giving the $L/M$ ratio as a function of distance from the region. The results of this analysis are presented in Fig.~\ref{rings}.

\begin{figure}
\includegraphics[scale=0.50]{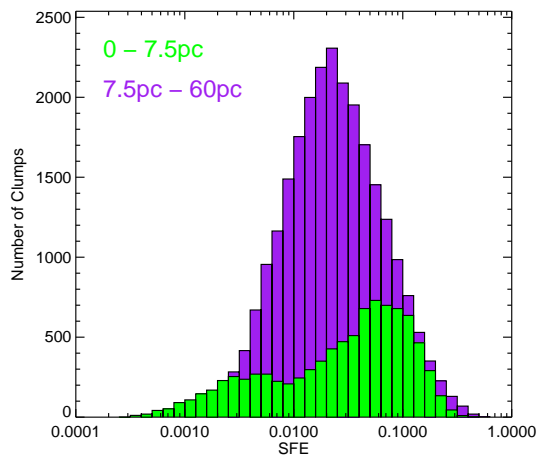}
\caption{Simulated SFEs for matched sources in W49A from Fig.~\ref{truecomp}, split by radius from the centre of W49A. Green bars represent sources within 7.5\,pc, with the purple bars representing sources between 7.5 and 60\,pc.}
\label{trueW49}
\end{figure}

The two regions have indistinguishable $L/M$ ratios in the inner three annuli, but W51 has significantly elevated $L/M$ at the outermost radii. The reason for this latter difference could be twofold. W49A is a relatively compact star-forming region, $\sim$\,20\,pc in the most extended direction, whereas W51 is much larger, extending to $\sim$\,40\,pc in one direction from the most compact part of the source (Figure \ref{W51map}).  The location of W51 in the tangent of the Sagittarius arm \citep{Sato10} may also contribute, since the outermost radii may include unassociated emission in the line of sight.

The dip in the central aperture of Fig.~\ref{rings} may be caused by depletion of the mass in the centre of the regions, the conditions may not be conducive to clump formation with potential sources broken up and therefore no further star formation, or the offset of star formation from the central regions of gas shells \citep{Thompson12,Palmeirim17}. However, as mentioned above, the central area of W51 contains saturated pixels, which may have prevented Hi-GAL source detections. To account for this for the purposes of this analysis, we consulted the ATLASGAL compact source catalogue \citep{Urquhart14b}, since the positional matching of Hi-GAL and ATLASGAL clumps is in good agreement in the W51 region, with $\sim$ 92 per cent of ATLASGAL sources corresponding to Hi-GAL sources. Four ATLASGAL clumps were found in the saturated regions and used as markers for possible Hi-GAL sources.  We then produced SEDs from the Hi-GAL image data using the method of \citet{Konig17}, with photometry within apertures of radii of 25\,arcsec, 1.5 times the median size of Hi-GAL sources associated with W51. The addition of this luminosity did not significantly alter the $L/M$ value in the central 7.5\,kpc of W51.

There is also evidence for radial structure in the probability distributions for the underlying SFE within clumps in W49A. If we split the population of clumps to examine sources in the first radial bin of Fig.~\ref{rings} with respect to the other three bins, the Monte-Carlo SFE simulation finds two very different distributions (Fig.~\ref{trueW49}). For clumps in the outer radial bins, the simulation finds a probability distribution that is very similar in form to that of W51 but with a peak at $\log({\rm SFE}) = -1.645\,\pm\,0.05$, corresponding to 2.2 per cent, which is significantly lower.  For the 10 clumps in the innermost radial region, it predicts a double-peaked distribution.  The lower SFE peak is associated with two high-mass, high-luminosity clumps in which the highest-mass stars can form, dominating the luminosity budget and limiting the fraction of clump mass converted into stars.  The remainder of the central subsample clumps tend to form mostly lower-mass stars, filling up the mass budget with relatively smaller contributions to the luminosity and producing the higher-SFE probability peak.  Such lowered SFEs in the highest-mass clumps is consistent with the prediction of \citet{Urquhart14b}. This suggestion of bimodality, hence mass segregation, echoes the hint of structure in the $L/M$ distribution in Fig.\ref{L/M} and is worthy of further investigation at higher resolution.

\subsection{Is the central region of W49A a mini-starburst?}

A number of regions in the Milky Way have been identified as potential mini-starburst regions, analogous to starburst galaxies in miniature.  Examples are RCW 106, Cygnus X, W43, W49, and W51 \citep{Rodgers60,Schneider06,NguyenLuong11,Galvan-Madrid13,Ginsburg15} with their inferred star-formation efficiency, the amount of star-forming material and current, ongoing star formation cited as reasons for this classification. However, the star-formation rate densities of W43, W49 and W51 are an order of magnitude greater than the other regions on this list, with W49 and W51 having an order of magnitude greater SFR than all other regions \citep{NguyenLuong16}. This, together with other results implying that the presence of W49 and W51 significantly affect the mean star-formation efficiency on kiloparsec scales \citep{Moore12}, makes it clear that these two regions are exceptional within the Galaxy. They also form part of the $\sim$ 30 complexes that contribute most of the star-formation rate and associated luminosity to the Milky Way \citep{Urquhart17}.

As the observable analogues of the star-formation efficiency are consistent with those in W51 and with other Galactic environments \citep[e.g.][]{Eden12} the cause of the starburst-like behaviour must be on larger scales than those confined to clumps. This points towards the ISM conditions within the whole of W49A as the source of the starburst-like conditions within this region.

Chemical analysis of the ISM in W49A has revealed starburst-like conditions within it (\citealt{Roberts11}; \citealt{Nagy12}). The high density and high temperature of the gas is comparable with the conditions found in ULIRGs \citep{Nagy15}.  The highest-temperatures ($\sim$\,200\,K) are preferentially tracing shocked regions, with these tracers found over a large area of W49A \citep{Nagy15}. However, we find the temperatures of the potentially star-forming clumps to be consistent with those found in other, more regular regions in the Galaxy, such as the W43 complex \citep{Eden12} and interarm sources \citep{Eden13}. The cool dust may be washing out the high temperature gas, due to a larger filling factor, as is potentially seen in external galaxies \citep{Watanabe17}.

An example of this is the formaldehyde (H$_{2}$CO) emission associated with a $3.3 \times 3.3$\,pc region in W49A and detected on kiloparsec scales in external starburst systems \citep{Mangum13}.  In W51, the formaldehyde emission and other assorted dense-gas tracers are associated with UC\ion{H}{ii} regions on scales of 0.1\,pc \citep{Zhang97}, whereas the large-scale H$_{2}$CO is observed in absorption \citep{Martin-Pintado85}.  Any future advancement in analysing the Galactic analogues of starburst conditions requires studying the chemical composition of the region \citep[e.g.][]{Nagy15}.

W49A is also rather unique in being a source of very high energy ($>100$\,GeV) $\gamma$-ray emission, as detected by the High Energy Stereoscopic System (HESS; \citealt{Brun11}), a phenomenon rare in Galactic star-forming regions and more commonly associated with starburst galaxies such as M82 and NGC253 \citep{Ohm12}.  Galactic sources are usually supernova remnants \citep{FermiColl17} and the mechanism is possibly fast proton collisions with dense gas producing $\pi^0$ decays \citep{Brun11}, such as in the W49B supernova remnant \citep{Keohane07}. However, W49A has two giant gas shells, with the shocks produced by the strong winds causing the $\gamma$-ray emission \citep{Peng10}.

\citet{Papadopoulos10} and \citet{Papadopoulos11} postulated that cosmic rays may be regulating the star formation in starburst systems, globally causing high molecular gas temperatures. ULIRGS are dominated by warm, dense gas \citep{Papadopoulos12}, conditions which could lead to a relatively top-heavy IMF \citep{Klessen07} by raising the effective Jeans mass. 

\section{Summary \& Conclusions}

We have compared the star-forming properties of W49A and W51, two major star-forming regions in the Milky Way whose presence affects the average properties of Galaxy-scale samples of young stellar objects and that are often referred to as Galactic starburst analogues.

We also present a new 850-$\upmu$m continuum map of a 1-degree diameter area around W49A, made using SCUBA-2 at JCMT, at a pixel-to-pixel rms of 17.39\,mJy\,beam$^{-1}$. 173 compact sources were extracted from this map using the {\sc FellWalker} \citep{Berry15} algorithm. By comparison with spectral line surveys, 61 of these were placed at the distance of W49A. 293 objects were found in the JCMT Plane Survey (JPS) compact-source catalogue \citep{Eden17} within a 60\,pc radius at the distance of W51.

The clump-mass distributions of the two regions are consistent with each other, having fitted power-law indices of $\alpha = -1.55 \pm 0.11$ and $\alpha = -1.51 \pm 0.06$. However, the luminosity distributions differ significantly, with W49A having a shallower fitted power-law index of $\alpha = -1.26 \pm 0.05$, compared to $\alpha = -1.51 \pm 0.03$ for W51. As the CMDs are consistent, but the LDs are not, this could be indicative of an underlying difference in the star-formation rate and efficiency in W49A. The flatter luminosity distribution, combined with elevated temperatures, high gas densities and the fact that W49A is a source of very high-energy $\gamma$-ray emission \citep{Brun11} suggest that it is the most promising candidate for a Galactic starburst analogue or mini-starburst.

The clump-formation efficiencies and $L/M$ ratios of the two regions are consistent with each other, as well as with other extreme star-forming regions in the Galaxy. The $L/M$ ratios and simulated SFEs found for the individual clumps within the two regions are also consistent with each other, except in the central regions of W49A, where the SFE probability distribution favours either low or high efficiencies within clumps.

\section*{Acknowledgements}

DJE is supported by a STFC postdoctoral grant (ST/M000966/1). This publication makes use of molecular line data from the Boston University-FCRAO Galactic Ring Survey (GRS). The GRS is a joint project of Boston University and Five College Radio Astronomy Observatory, funded by the National Science Foundation under grants AST-9800334, AST-0098562, \& AST-0100793. This work is part of the VIALACTEA Project, a Collaborative Project under Framework Programme 7 of the European Union, funded under Contract \#607380 that is hereby acknowledged. PACS has been developed by a consortium of institutes led by MPE (Germany) and including UVIE (Austria); KU Leuven, CSL, IMEC (Belgium); CEA, LAM (France); MPIA (Germany); INAF-IFSI/OAA/OAP/OAT, LENS, SISSA (Italy); IAC (Spain). This development has been supported by the funding agencies BMVIT (Austria), ESA-PRODEX (Belgium), CEA/CNES (France), DLR (Germany), ASI/INAF (Italy), and CICYT/MCYT (Spain). SPIRE has been developed by a consortium of institutes led by Cardiff University (UK) and including Univ. Lethbridge (Canada); NAOC (China); CEA, LAM (France); IFSI, Univ. Padua (Italy); IAC (Spain); Stockholm Observatory (Sweden); Imperial College London, RAL, UCL-MSSL, UKATC, Univ. Sussex (UK); and Caltech, JPL, NHSC, Univ. Colorado (USA). This development has been supported by national funding agencies: CSA (Canada); NAOC (China); CEA, CNES, CNRS (France); ASI (Italy); MCINN (Spain); SNSB (Sweden); STFC, UKSA (UK); and NASA (USA). This research has made use of NASA's Astrophysics Data System. The JCMT has historically been operated by the Joint Astronomy Centre on behalf of the Science and Technology Facilities Council of the United Kingdom, the National Research Council of Canada and the Netherlands Organization for Scientific Research. Additional funds for the construction of SCUBA-2 were provided by the Canada Foundation for Innovation. This research has made use of NASA's Astrophysics Data System. The Starlink software \citep{Currie14} is currently supported by the East Asian Observatory. DJE would like to dedicate this work to his uncle, Joseph Eden.

\bibliographystyle{mnras}
\bibliography{W49W51_ref}

\appendix

\section{W51 JCMT Plane Survey Data}
\label{W51app}

\begin{figure*}
\includegraphics[width=0.99\linewidth]{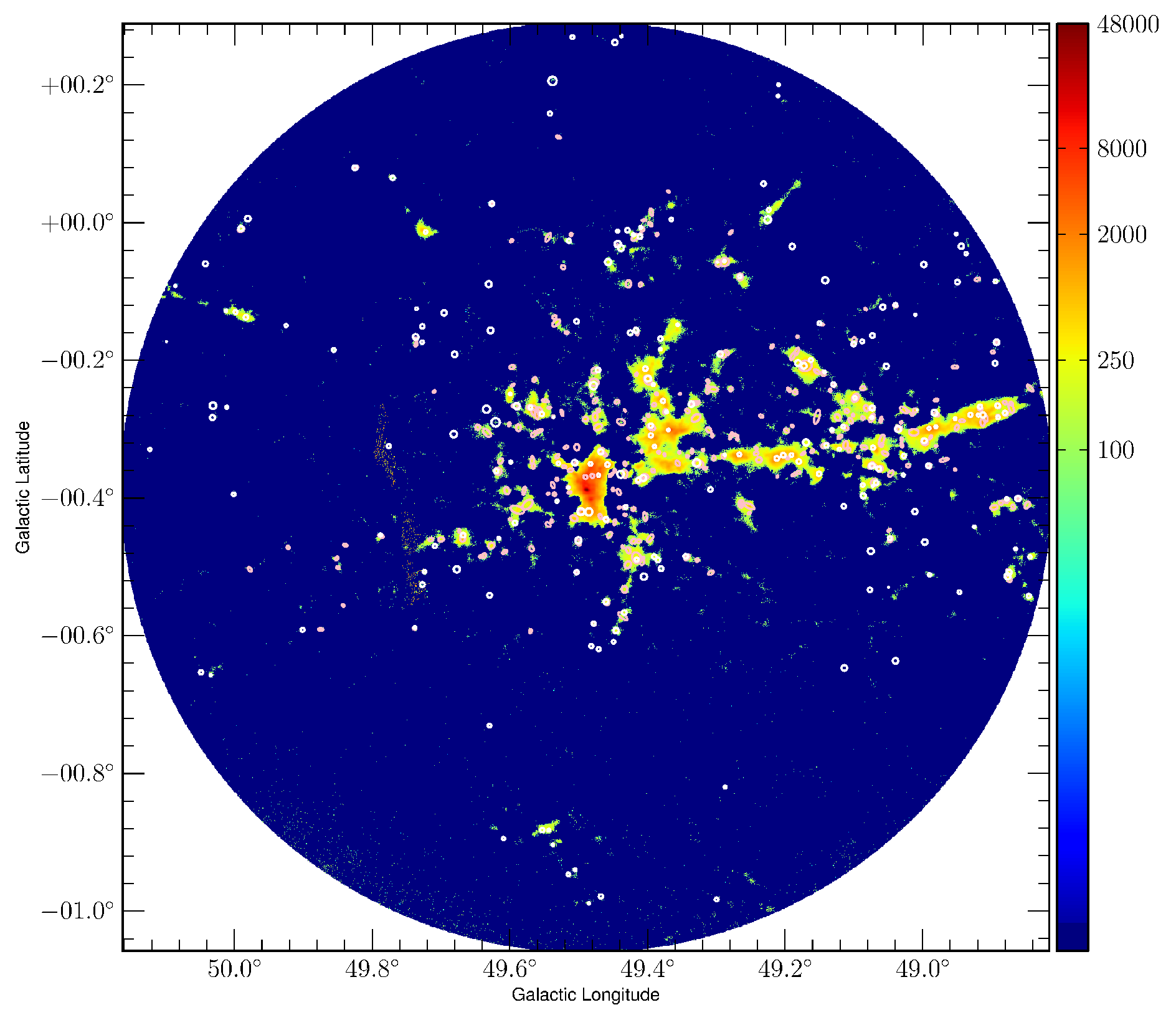}
\caption{The JCMT Plane Survey map of the W51 region. The intensity scale is in units of mJy\,beam$^{-1}$. The entire 60\,pc radius is included in this image. The dynamic range is scaled in such a way that only pixels above 3$\upsigma$ are displayed, however, there is a patch of noise at $\ell \simeq 49.75^{\circ}$ that could not be removed in the reduction process. The pink ellipses represent the JCMT sources assigned to W51, with the small white circles showing the positions of the Hi-GAL sources.}
\label{W51map}
\end{figure*}

The data used for the W51 analysis is part of the JCMT Plane Survey (JPS; \citealt{Moore15,Eden17}), specifically the $\ell$\,=\,50$\degr$ field. The JPS data is used within a 40-arcminute radius centred on $\emph{l}$ = 49$\fdg$486, $\emph{b}$ = -0$\fdg$381, which corresponds to a radius of 60\,pc. The image is presented in Fig.~\ref{W51map}.

\section{Hi-GAL and JPS Sources}
\label{sourceIDs}

The source IDs and positions of the JPS sources and the Hi-GAL sources used for the W51 CMD and the two LDs are listed in Tables~\ref{JPSsources}, \ref{W49HiGAL}, and \ref{W51HiGAl}. The complete versions of these tables can be found in the Supporting Information to the online article.

\begin{table}
\begin{center}
\caption{The JPS sources associated with the W51 star-forming region. A complete version of this table containing the 293 sources can be found in the Supporting Information to the online article.}
\label{JPSsources}
\begin{tabular}{lcc}\hline
JPS Source & $\ell_{\rmn{peak}}$ & $\emph{b}_{\rmn{peak}}$ \\
ID & ($^{\circ}$) & ($^{\circ}$) \\
\hline
JPSG049.490-00.370	&	49.490	&	-0.369	\\
JPSG049.480-00.368	&	49.480	&	-0.367	\\
JPSG049.787-00.292	&	49.787	&	-0.292	\\
JPSG048.913-00.290	&	48.913	&	-0.290	\\
JPSG049.751-00.471	&	49.750	&	-0.471	\\
JPSG049.768-00.356	&	49.768	&	-0.356	\\
JPSG048.908-00.285	&	48.908	&	-0.284	\\
JPSG048.997-00.319	&	48.997	&	-0.318	\\
JPSG049.495-00.420	&	49.495	&	-0.420	\\
JPSG049.561-00.280	&	49.561	&	-0.279	\\
\hline
\end{tabular}
\end{center}
\end{table}

\begin{table}
\begin{center}
\caption{The Hi-GAL sources associated with the W49A star-forming region. A complete version of this table containing the 57 sources can be found in the Supporting Information to the online article.}
\label{W49HiGAL}
\begin{tabular}{lcc}\hline
Hi-GAL Source & $\ell_{\rmn{peak}}$ & $\emph{b}_{\rmn{peak}}$ \\
ID & ($^{\circ}$) & ($^{\circ}$) \\
\hline
180502	&	42.856	&	-0.112	\\
180640	&	42.899	&	-0.061	\\
180663	&	42.907	&	-0.005	\\
180674	&	42.910	&	-0.025	\\
180724	&	42.925	&	-0.069	\\
180732	&	42.927	&	-0.054	\\
180740	&	42.930	&	-0.041	\\
180746	&	42.933	&	-0.016	\\
180756	&	42.937	&	-0.068	\\
180765	&	42.942	&	-0.035	\\
\hline
\end{tabular}
\end{center}
\end{table}

\begin{table}
\begin{center}
\caption{The Hi-GAL sources associated with the W51 star-forming region. A complete version of this table containing the 434 sources can be found in the Supporting Information to the online article.}
\label{W51HiGAL}
\begin{tabular}{lcc}\hline
Hi-GAL Source & $\ell_{\rmn{peak}}$ & $\emph{b}_{\rmn{peak}}$ \\
ID & ($^{\circ}$) & ($^{\circ}$) \\
\hline
196057	&	48.839	&	-0.438	\\
196078	&	48.846	&	-0.240	\\
196092	&	48.850	&	-0.409	\\
196135	&	48.861	&	-0.401	\\
196164	&	48.869	&	-0.414	\\
196172	&	48.871	&	-0.269	\\
196184	&	48.875	&	-0.508	\\
196191	&	48.876	&	-0.256	\\
196193	&	48.877	&	-0.514	\\
196197	&	48.878	&	-0.401	\\
\hline
\end{tabular}
\end{center}
\end{table}

\bsp
\label{lastpage}

\end{document}